# Enzymatic AND-Gate Based on Electrode-Immobilized Glucose-6-Phosphate Dehydrogenase: Towards Digital Biosensors and Biochemical Logic Systems with Low Noise

**Vladimir Privman**,[†]  **Valber Pedrosa**,[‡]  **Dmitriy Melnikov**,[†]

**Marcos Pita**,[†]  **Aleksandr Simonian**[‡]  and  **Evgeny Katz**[†*]

[†] Department of Chemistry and Biomolecular Science, and
Department of Physics, Clarkson University, Potsdam, NY 13699, USA

[‡] Materials Research and Education Center, Auburn University, Auburn, AL 36849, USA

**Abstract**

Electrode-immobilized glucose-6-phosphate dehydrogenase is used to catalyze an enzymatic reaction which carries out the **AND** logic gate. This logic function is considered here in the context of biocatalytic processes utilized for the biocomputing applications for "digital" (threshold) sensing/actuation. We outline the response functions desirable for such applications and report the first experimental realization of a sigmoid-shape response in one of the inputs. A kinetic model is developed and utilized to evaluate the extent to which the experimentally realized gate is close to optimal.

Click this area for *future updates* of this article and for the file with *higher-resolution* images.

[*] Corresponding author: phone +1-315-268-4421, e-mail ekatz@clarkson.edu

# 1. Introduction

Recently there has been significant interest in chemical (Credi, 2007; De Silva and Uchiyama, 2007; Pischel, 2007; Szacilowski, 2008) and biochemical (Shao et al., 2002; Ezziane, 2006; Unger and Moult, 2006) information processing, including that based on enzyme reactions (Sivan et al., 2003). Enzymatic reactions have been developed to mimic digital logic gate functions (Baron et al., 2006a; Strack et al., 2008a) and certain elementary arithmetic operations (Baron et al., 2006b). Biochemical reactions "networked" in a Boolean logic circuit (Niazov et al., 2006; Strack et al., 2008b), promise new applications such as multiple-input sensing resulting in response/actuation of the "digital" (threshold) nature (Manesh et al., 2009; Pita et al., 2009). For example, the output chemical concentration reaching a certain logic-**1** value, could signal that an action is needed, whereas concentrations at logic-**0** (typically, at 0) would be the "no action needed" value.

Here we address interesting aspects of the difference between the usual "proportional" (linear response) sensor functioning and response properties desirable in new digital biochemical logic designs. We consider the **AND** logic gate function based on electrode-immobilized enzyme glucose-6-phosphate dehydrogenase (G6PDH). The biocatalytic reaction

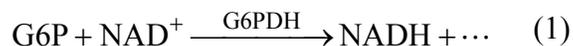
$$\text{G6P} + \text{NAD}^+ \xrightarrow{\text{G6PDH}} \text{NADH} + \cdots \quad (1)$$

is studied, with two inputs: glucose-6 phosphate (G6P) and cofactor nicotinamide adenine dinucleotide ($NAD^+$), and one output: the reduced cofactor (NADH). The other reaction product, not shown in (1), is 6-phosphogluconate. The output signal is detected electrochemically. It should be noted that G6PDH is used in various biosensors (Cui et al., 2007; Tzang et al., 2001), including the analysis of ATP activity (Cui et al., 2008), which has important biomedical applications (Tanvir et al., 2009).

To facilitate the discussion, let us for the moment consider response curves for the pass-through (identity) function which corresponds to the ordinary, proportional sensing with one input being "sensed" by one output: This is shown in Figure 1: curve (a), as well as (b), where for the latter the input and output were rescaled in terms of their reference values for the digital logic-**0** (assumed to be the physical zero) and logic-**1** (determined by the application). These two shapes illustrate the concave and convex response functions; the shaded areas mark the approximately linear response appropriate for proportional sensing. The convex shape is typical for (bio)chemical systems because they usually reach saturation due to limited reaction rates and

– 2 –

reactant supply. However, the desirable response for digital logic gates and circuits should be sigmoid, illustrated by curve (c). Indeed, this response decreases small "analog" noise around the logic-**0** and **1** values, as discussed in detail in our earlier work (Privman et al., 2008).

While such a sigmoid response is observed in natural systems (Setty et al., 2003; Buchler et al., 2005; Alon, 2007), it is generally not easily achieved in biochemical enzyme-based reactions. Furthermore, additional issues arise for two-input gate functions, such as the one studied here which at the digital **0,0**; **0,1**; **1,0**; **1,1** inputs yields the **AND** output: **1** for input **1,1**; **0** otherwise. We report that the process in (1), carried out with electrode-immobilized enzyme, experimentally displays sigmoid behavior in one of its inputs: G6P. Section 2 provides the experimental information. After describing our experimental results in Section 3.1, we will develop, in Section 3.2, a model suitable for the biochemical reaction studied and for the **AND** gate that it realizes. After formulation of the kinetics for the biochemical system in Section 3.3, we present our results for the fitting of the experimental data in Section 3.4. Section 3.5 offers further results and summary.

2. Experimental

*2.1. Chemicals and reagents*

The chemicals were purchased from Sigma-Aldrich and used as supplied without further purification: glucose-6-phosphate dehydrogenase (G6PDH) from *Leuconostoc mesenteroides* (E.C. 1.1.1.49), β-nicotinamide adenine dinucleotide (NAD$^+$), D-glucose-6 phosphate anhydrous (G6P), other salts and reagents of analytical quality. Multiwalled carbon nanotubes (MWCNTs) (purity 95%, length 1-5 μm, diameter 30±10 nm) were purchased from Nanolabs (Newton, MA) and purified according to the procedure described below. Ultrapure water (18.2 MΩ cm) from Millipore Direct-Q source was used in all of the experiments.

*2.2. Electrode modification*

A suspension of 1 mg of MWCNTs was sonicated at room temperature for 10 hours in a mixture of 3:1 H$_2$SO$_4$ (conc.) and HNO$_3$ (conc.). The contents were cooled to ambient temperature and MWCNTs allowed settling to the bottom. The MWCNTs were extensively washed with water and filtered by centrifugation until the pH of the solution was neutral.

The glassy carbon (GC) disk electrode (BASi, USA) with a diameter of 3 mm was polished prior to its modification with 0.3 and 0.05 µm alumina slurry (MicropolishII, Buehler, IL, USA) and then sonicated in ethanol for 10 min. Then the GC electrode was modified by deposition of a 4.0 µL droplet of MWCNTs obtained in the previous purification step and dried in air. The next modifying solution was prepared by mixing 20 µL of G6PDH (60 units), 20 µL of glutaraldehyde, 1% v/v in phosphate buffered saline (PBS) (PBS = 137 mM NaCl, 2.7 mM KCl, 10 mM $Na_2HPO_4$ and 1.76 mM $KH_2PO_4$) pH = 7.4, 1.0% (w/v) of BSA, and 2 µL of Nafion solution, 0.5% w/v. Then 10 µL of the modifying solution were deposited on the surface of the working GC electrode, followed by drying and storing at 4°C overnight. After that, the electrode was rinsed with PBS buffer and then ready for electrochemical measurements.

## 2.3. Instruments and measurements

The electrochemical measurements were performed using a CH Instrument Model CHI630C with an electrochemical cell consisting of the enzyme-modified working electrode, Ag/AgCl (3 M NaCl) reference and a platinum wire counter electrode. All cyclic voltammograms were obtained with the potential scan rate of 30 mV $s^{-1}$ in the electrolyte solution composed of 10 mM PBS buffer (pH = 7.4) in the presence of 10 mM $MgCl_2$ at ambient temperature (23 ± 2)°C. The experiments were carried out by potential scanning between 0.6 and −0.3 V for 10 cycles until a steady cyclic voltammogram was obtained, under stirring condition.

## 3. Results and Discussion

### 3.1. Cyclic voltammetry measurements of the enzyme-modified electrode reaction

The activity of the G6PDH-modified electrode was followed by cyclic voltammetry measurements in the presence of variable concentrations of G6P (substrate) and $NAD^+$ (cofactor) in the solution. The biocatalytic reaction of the immobilized enzyme, (1), resulted in the production of the reduced cofactor (NADH), which was re-oxidized electrochemically. The oxidation of NADH was observed as the anodic peak on the cyclic voltammograms at $E_{\text{pa}} = 0.26$ V, Figure 2, curve (a). In the absence of one or both of the reactants (G6P and NAD$^+$), the cyclic voltammograms did not show any peak at this potential, curve (b). The peak current, $\Delta I$, was considered as the output signal of the enzyme-based **AND** logic gate. The $\Delta I$ value was defined as the difference between the maximum current value and the interpolated baseline, as illustrated in Figure 2 for curve (a).

From the cyclic voltammograms, one can conclude that NADH can be directly oxidized at the modified GC electrode. The set of cyclic voltammograms obtained at variable concentrations of both inputs, G6P and NAD$^+$, was used to map the response surface of the enzyme-modified electrode. The concentration ranges for both chemical inputs, G6P and NAD$^+$, were selected to demonstrate the sigmoid domain of the output signal function, as presented below. The value of the measured output signal, $\Delta I$, was converted to the effective NADH concentration near the electrode surface according to the relation $\Delta I/[\text{NADH}] = 19\,\mu\text{A/mM}$, established by calibration measurements performed in the presence of standard additions of NADH (not shown).

## 3.2. Gate-function response surface

The digital values of the chemical concentrations correspond to 0, for logic-**0**. For logic-**1**, the values are set by the gate's environment or by its networking with other reactions, for the two input concentrations at time $t = 0$, as well as the gate-function-determined value of the output concentration at the time of the voltammogram taking, which is actually a short interval but will be collectively denoted as $t = t_{\text{gate}}$,

$$[\text{NAD}^+]_{\max} = [\text{NAD}^+]_{\text{logic-1}}(t=0), \qquad (2)$$

$$[\text{G6P}]_{\max} = [\text{G6P}]_{\text{logic-1}}(t=0), \qquad (3)$$

$$[\text{NADH}]_{\max} = [\text{NADH}]_{\text{logic-1}}(t=t_{\text{gate}}). \qquad (4)$$

As described in recent works (Privman et al., 2008; 2009), consideration of networking of a gate as part of a larger "circuit" for biochemical logic applications, requires control of the buildup of noise. The *level* of the noise per se largely depends on the gate's environment. Rather it is the degree of noise *amplification* that should be kept in check to ensure the stable, scalable operation of increasingly complex networks. In fact, for small to moderate size networks, control of the analog noise amplification should be considered, as accomplished by the appropriate gate response function shape alluded to in Figure 1. For larger networks, "digital" noise buildup becomes dominant and should also be controlled. The latter is usually accomplished by network design and requires additional network elements such as filters and signal splitters. These issues have been discussed in detail, Privman et al., 2008; 2009, and will not be reviewed here.

The key observation is that for a single gate, one has to map out its response for input and output values not only at the logic-**0** and **1** but also for the interval from 0 to 1 (and possibly for somewhat larger values, past 1) of the reduced concentration variables defined as

$$x = [\text{NAD}^+](t=0)/[\text{NAD}^+]_{\max}, \qquad (5)$$

$$y = [\text{G6P}](t=0)/[\text{G6P}]_{\max}, \qquad (6)$$

$$z = [\text{NADH}](t=t_{\text{gate}})/[\text{NADH}]_{\max}. \qquad (7)$$

Thus, we scan inputs between 0 and the maximal, logic-**1** values, and record the corresponding output, to map out the gate-function response surface,

$$z = F(x, y). \qquad (8)$$

The actual data are presented and analyzed later.

As with the single-input gate-function response curves normalized to the intervals from 0 to 1 in both variables, see Figure 1, for the normalized *two-input* response surface the functional form at the logic values corresponds to the Boolean results, here, $F(1,1) = 1, F(0,0) = 0, F(0,1) = 0, F(1,0) = 0$ of the **AND** gate. However, due to noise the inputs and the output can also be spread in the vicinity of the logic values. A typical response of biocatalytic two-input reactions is shown in Figure 3, surface (a). It is smooth near all the logic points. Therefore, the noise spread amplification for small noise levels can be estimated by the absolute values of the gradients, $|\vec{\nabla}F|_{00,01,10,11}$, of the function $F$ at the logic points. The parameters of the gate should be such that the largest of these four gradients — which literally measures the noise amplification factor — is not much exceeding 1. However, as described in earlier work (Privman et al., 2008; 2009; Melnikov et al., 2009), such an optimization of the gate function is usually difficult or impractical because it requires large changes, but orders of magnitude, of the "gate machinery" parameters, such as the enzyme concentration. Moreover, the noise amplification factor cannot be much less than 1.2 for such gates (Privman et al., 2008; 2009; Melnikov et al., 2009), so that at most order 10 of them can be networked before additional resource consuming filtering or "digital" noise reduction techniques are required.

Ideally, we would like to have the response shown in Figure 3, surface (b), which generalizes the "sigmoid" property of Figure 1, curve (c) and therefore can have small gradients at *all* of the logic points. With the largest gradient less than 1, such a gate would actually offer analog noise spread *reduction*, i.e., incorporate a filtering feature in its function. While such functions are encountered in some natural processes (Setty et al., 2003; Buchler et al., 2005; Alon, 2007), they have been allusive in simple (bio)chemical reactions and have not thus far been realized experimentally.

Given the afore-described difficulties with gate optimization, other options should be considered. For instance, the case in Figure 3, surface (c) illustrates a gate response (Melnikov et al., 2009) which has a rounded "ridge." While the slopes along the ridge are not that small, the larger slopes are limited in their directional spread: most of the surface (c) is nearly planar. Therefore, when the noise *distribution* is carefully analyzed (Melnikov et al., 2009; see also Section 3.5 below), one finds that with a proper adjustment of parameters within reasonable ranges of variation, the noise amplification factor can be actually reduced to close to 1 (means, practically no noise amplification). Another approach (Privman et al., 2009) with non-sigmoid gates is to rely on *network* optimization.

Here we consider an interesting option: a response "sigmoid" in only one of the two inputs. An illustration is offered in Figure 3, surface (d). Note that this plot is just a schematic. Our actual data are presented in Figure 4 and they are "sigmoid" in the variable $y$, rather than in $x$, though the choice of which input to call $x$ or $y$ is arbitrary, determined entirely by convenience in applications. We argue that while such functions cannot offer noise reduction, they can definitely be optimized, with parameter adjustment within reasonable ranges, to yield practically *no noise amplification*, i.e., have the largest gradient close to 1.

### *3.3. Rate equations for the biocatalytic reaction*

The biocatalytic process in (1), is definitely not a direct reaction. It involves several steps and pathways (Soldin and Balinsky, 1968; Shreve and Levy, 1980; Özer et al., 2001), some reversible, the precise kinetics of which is not fully understood, especially for the electrode-immobilized enzyme as the biocatalyst. We found the output of this reaction sigmoid as a function of the concentration of the substrate, G6P, as illustrated in Figure 4. Note that such a property, while useful for "Boolean-type sensing," actually constitutes an impediment to possible utilizations of this enzymatic system in "proportional sensing" applications which ordinarily require a well defined linear regime illustrated in Figure 1, curve (b).

Two comments are in order before we formulate the rate equations to be used. As mentioned, the actual kinetics of the processes involved is not fully understood. Specifically, there have been indications (Scopes, 1997) that G6PDH can have allosteric properties, but under conditions and for enzyme from sources different from ours. A sigmoid response of G6PDH enzyme similar (but not identical) to one used in our work was reported (Anderson et al., 1997). Thus, we don't actually know if the sigmoid behavior suggested by Figure 4 is due to the property of the enzyme or caused by some electrode-related processes including possibly enzyme restructuring. While investigation of this issue is interesting from the point of view of enzymatic kinetics, for our logic-gate parameterization we can seek a simplified, few-parameter description, which will bypath the issues of the kinetic origin of the experimentally observed "self-promoter" property of G6P.

Thus, we use a phenomenological rate equation approach which is introduced here with two goals in mind. We would like to have a "generic" parameterization which can be used for other (bio)chemical systems for which a self-promoter property is observed with respect to any of the inputs. While simplified, the rate equations should capture the dependence on the key parameters that can be adjusted to control the gate performance. To further explain this, let us write down the rate equation that will be used for the concentration of G6P, denoted by $[G6P](t) = G(t)$ for brevity,

$$\frac{dG}{dt} = -[\alpha + \beta(G_0 - G)]GM . \qquad (9)$$

We assume that the substrate is captured by the enzyme, the concentration of which is $[G6PDH](t) = M(t)$, and we ignore the possible irreversibility of this reaction step, which is confirmed a posteriori for our regime because the equations fit the data, see Section 3.4, though the main reason for the latter assumption is to minimize the number of adjustable parameters in our data fitting. The reaction rate is initially proportional to $\alpha G(t)M(t)$, while the "self-promoter" property is included by the following modification: for later times the rate increases proportionately to the amount of the consumed substrate, $\alpha \to \alpha + \beta[G_0 - G(t)]$, where $G_0 = G(0)$ is defined as the initial concentration of the input G6P.

Note that the rate equation (9), as well as the other rate equations to be presented shortly, contain several parameters that can be adjusted to try to optimize the gate functioning. Specifically, we could change the reaction time $t_{\text{gate}}$ at which the output is recorded. While the (initial) input concentrations such as $G_0$ are likely fixed, as discussed earlier, we can adjust the "gate machinery" by changing the initial amount (activity) of the enzyme $M_0 = M(0)$ in many cases, though in our system the concentration (activity) of the catalytically active enzyme at the electrode is not easy to estimate and it de-facto becomes one of the fitting parameters; see Section 3.4. The rate constant parameters $\alpha$, $\beta$, etc., can also be adjusted, though their variation, requiring changes in the physical or chemical conditions of the reaction, is not always easy to predict for enzymatic systems.

The substrate-capture step, (9) results in a complex, of concentration $C(t)$. The latter in turn combines with $NAD^+$, of concentration $[NAD^+](t) = N(t)$, to yield the product, of concentration $[NADH](t) = P(t)$, as well as restoring the biocatalyst. We will use this rather simplified description to write the remaining rate equations for our system,

$$\frac{dC}{dt} = [\alpha + \beta(G_0 - G)]GM - \gamma NC, \quad (10)$$

$$\frac{dM}{dt} = -[\alpha + \beta(G_0 - G)]GM + \gamma NC, \quad (11)$$

$$\frac{dN}{dt} = -\gamma NC, \quad (12)$$

$$\frac{dP}{dt} = \gamma NC, \quad (13)$$

where $\gamma$ is another rate-related parameter for the data fitting. The initial conditions were partly specified earlier, and the remaining ones are $C(0) = 0$, $P(0) = 0$, $N_0 = N(0) = [NAD^+](t = 0)$. These rate equations will be used to fit our data, as presented in the next subsection, as well as to explore the dependence on the parameters, in order to estimate the noise-amplification properties of our biocatalytic system when used as an **AND** logic gate.

*3.4. Experimental data fitting and gate-function properties*

The experimental response surface is shown in Figure 4, in which the concentration variables are not scaled to the "logic" values; the fitting of these data with the rate equations (9)-(13) is also shown. The experimental points were obtained from the measured current by using the conversion factor of $\Delta I / [\text{NADH}] = 19\,\mu\text{A/mM}$, determined from the NADH calibration as mentioned earlier. Since the experimental response function is rather noisy, the fitting was performed as a weighted non-linear least-squares fit of the data with emphasis in the region of small [G6P], for which the self-promoter property is observed. This yielded estimates $\alpha = 0.03\,\text{mM}^{-1}\text{s}^{-1}$, $\beta = 42\,\text{mM}^{-2}\text{s}^{-1}$, $\gamma = 1.05\,\text{mM}^{-1}\text{s}^{-1}$.

We point out that in addition to the three rate parameters, an effective initial enzyme concentration, $M_0$, can also be regarded as another not well-determined parameter. The applied enzyme is only partially immobilized in the thin film cross-linked at the electrode surface and the unbound enzyme is washed out, giving unknown load of the enzyme at the electrode interface, which is usual for this kind of immobilization procedures. The active enzyme concentration de-facto becomes another fitting parameter, for which we used the effective value $M_0 = 1\,\mu\text{M}$. However, the sensitivity of the data fitting to the inclusion of this concentration as another adjustable parameter is weak. Therefore, its value remains uncertain up to a factor of 2 times smaller or larger: the other parameters adjust to result in no substantial changes in the data fitting quality or in the values of the gradients at the logic points.

With our estimates for the reaction parameters, we thus computed the gradients at the four logic points as functions of the effective enzyme concentration and reaction time. The maximum of these gradients in the parameter range covered, happens to be at the logic-**0,1** point: $|\vec{\nabla}F|_{01}$. Interestingly, our randomly selected (for experimental convenience) values of the parameters yield the noise amplification factor 1.16, which is already better that those potentially achievable (but requiring dramatic variation of parameter values) in earlier studied enzymatic systems (e.g., Privman et al., 2008), with typical, smooth response surfaces of the type shown in Figure 3, surface (a). Within our model, we can seek a regime to "optimize" the gate "machinery" by changing the enzyme activity and/or the process (gate) time, while keeping all the other parameters fixed. This is shown in Figure 5: There is a broad region corresponding to reaction times longer than ~ 200 s and to enzyme activities comparable or larger than in our system, which yield practically no noise amplification, i.e., the largest gradient somewhat larger than but very close to 1, with values as low as ~1.05 within experimental reach.

*3.5. Additional results and concluding discussion*

The estimates in Section 3.4 based on the response-function gradients, correspond to very low level of noise, appropriate in electronics. However, in chemical systems noise at the level of a couple of percent of the maximum, logic-**1** reference value should be considered, for possible fluctuations away from both the logic-**0** and -**1** reference values. Thus, following earlier work (Privman et al., 2008; Melnikov et al., 2009) we also estimated noise amplification by the gate by calculating the width of the output signal distribution, $\sigma_{\text{out}} = \sigma_z$, as a function of the width of the input noise distributions, $\sigma_{\text{in}} = \sigma_{x,y}$, which for simplicity will be assumed equal for both inputs. We varied the value of $\sigma_{\text{in}}$ in the range from 0 to 30%, as shown in Figure 6, where we consider the largest of the output distribution spreads among the four logic inputs,

$$\sigma_{\text{out}}^{\max} = \max\left(\sigma_{\text{out}}^{00}, \sigma_{\text{out}}^{01}, \sigma_{\text{out}}^{10}, \sigma_{\text{out}}^{11}\right). \qquad (14)$$

Furthermore, we assumed uncorrelated, Gaussian (Privman et al., 2008) input noise distributions (half Gaussian for $x, y \geq 0$ at logic-**0**). The output noise distribution width $\sigma_{\text{out}}$ is then estimated as

$$\text{logic-}\mathbf{0}: \qquad \sigma_{\text{out}}^2 = \langle z^2 \rangle, \qquad (15)$$

$$\text{logic-}\mathbf{1}: \qquad \sigma_{\text{out}}^2 = \langle z^2 \rangle - \langle z \rangle^2, \qquad (16)$$

for each logic input, where the moments such as $\langle z^2 \rangle$ of the gate response function $z = F(x, y)$ were computed with respect to the input noise distributions (see Privman et al., 2008; Melnikov et al., 2009, for details). This gives the spread of the output signal near the four logic inputs. In general one would want to have maximum of these spreads, $\sigma_{\text{out}}^{\max}$, to be as small as possible. In fact, for network scalability the actual value of the noise spread is probably not as important as the degree of noise amplification by the gate, given by the "quality measure" $\sigma_{\text{out}}^{\max} / \sigma_{\text{in}}$; see Figure 6.

In Figure 6, curve (a) corresponds to our experimental conditions, i. e., the effective (active) enzyme concentration 1 μM and the reaction time 300 s. We compared this calculation with the one for the enzyme concentration 2 μM and time 400 s, for which the calculated maximum gradient was ~1.05; see Figure 5. The latter parameter values yield curve (b) in Figure 6. In the limit of small noise levels, $\sigma_{in} \to 0$, the ratio $\sigma_{out}^{max}/\sigma_{in}$ approaches the largest gradient value, the one at the logic-**0,1** point, estimated in Section 3.4, which is lower for curve (b). However, the value $\sigma_{out}^{10}$ overtakes the $\sigma_{out}^{01}$ as $\sigma_{out}^{max}$ for somewhat larger noise levels, at the points with the discontinuous slope (see Figure 6). Therefore, as seen from Figure 6, for input noise levels exceeding ~ 0.8%, the advantage of changing the experimental parameters, curve (a), to those corresponding to curve (b), is rapidly lost, because the "quality measure" $\sigma_{out}^{max}/\sigma_{in}$ can be kept closer to one, and in fact under ~ 1.2, for noise levels up to about 5% for our original experimental parameter values.

For small noise levels, the logic-**1,0** point on the "sigmoid side" of the response function, Figure 4, yields small output noise, as expected. However, when the fluctuations in the inputs increase, this logic point in fact becomes the "noisiest" of all four. This is because at larger concentrations this side of the gate function can actually be rather steep (see Figure 4), and since the sigmoid part is limited to a very small interval of concentrations close to zero, its presence plays no role at larger input fluctuations away from 0. In general, not only is the mere presence of the sigmoidality in the response function important for avoiding noise amplification, but the positioning of the inflection region is as well: For optimal gate-function parameters, the inflection should be at a concentration well away from both the logic-**0** and -**1** reference values.

In summary, we reported the first realization and performance analysis of an enzymatic **AND** gate with a noise-reducing "sigmoid" response to one of its two inputs. The studied enzymatic reaction was found to have a relatively small degree of analog noise amplification with the initially selected experimental parameters, which turned out to be in the theoretically calculated optimal regime.

**Acknowledgements**

The team at Clarkson University acknowledges support by the National Science Foundation (grants CCF-0726698, DMR-0706209), by Office of Naval Research (award #N00014-08-1-1202), and by the Semiconductor Research Corporation (award 2008-RJ-1839G).

The team at Auburn University acknowledges support from the USDA-CSREES (grant 2006-34394-16953). Additionally, this material is based on the work (A.S.) which was supported by the National Science Foundation, while working (A.S.) at the Foundation.

Any opinion, finding, and conclusions or recommendations expressed in this material are those of the authors and do not necessarily reflect the views of the National Science Foundation, the Department of Defense, or the U.S. Government.

# Figures

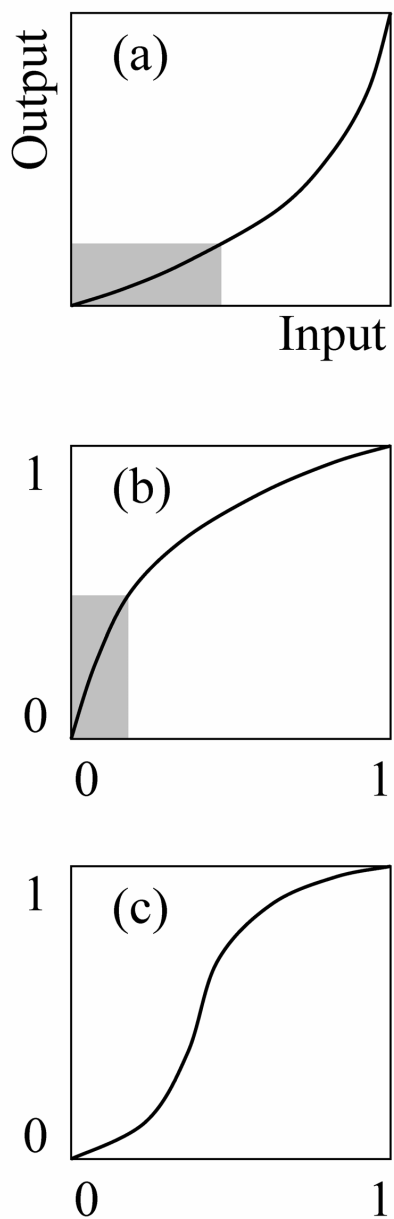

**Figure 1:** Schematic of single-input response curves shapes. (a) The concave response curve. The shaded area marks the approximately linear behavior. (b) The convex response, typical of (bio)chemical systems, with the axes rescaled to the logic-**0**,**1** reference values and the shaded area again marking the linear region. (c) The sigmoid response. The latter is desirable for digital logic applications even though it has a relatively small linear region (not marked), or no linear region at all if the slope at the origin is zero.

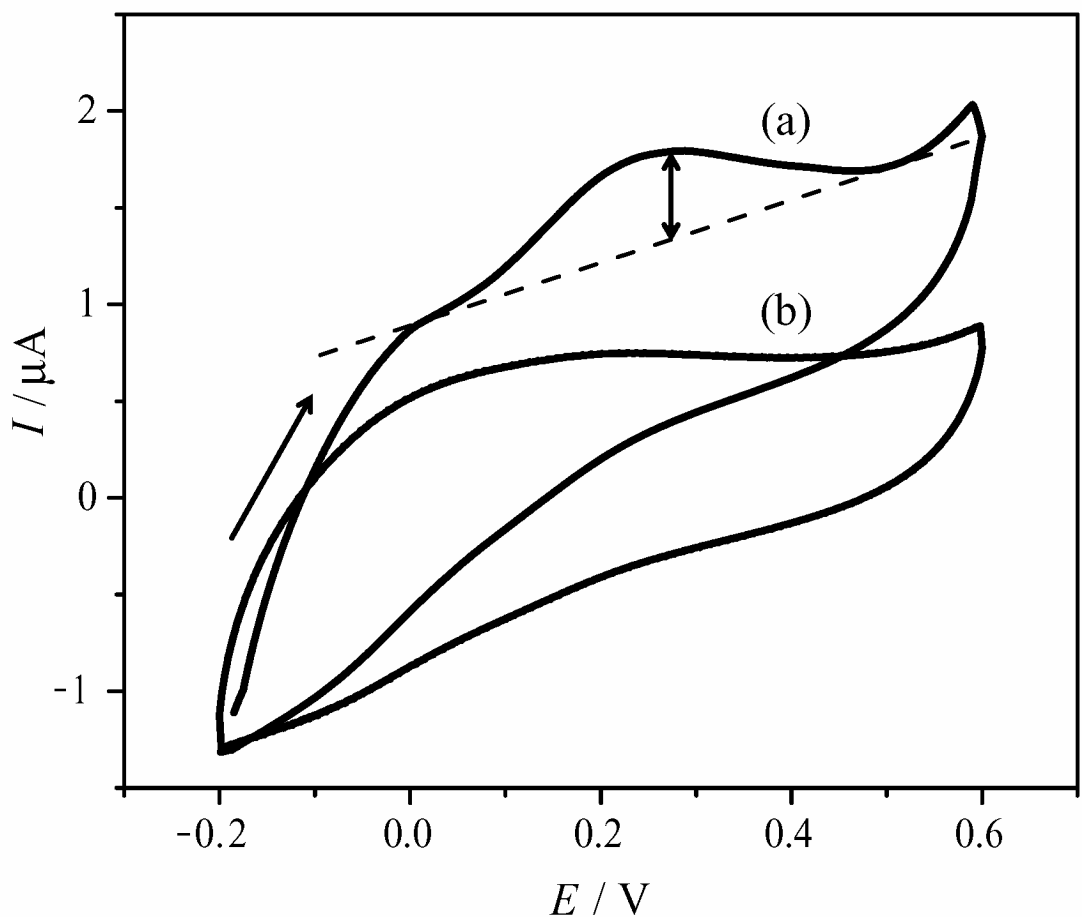

**Figure 2:** Sample cyclic voltammograms of the G6PDH-modified electrode obtained for different combinations of the input signals: (a) in the presence of G6P, 500 μM, NAD$^+$, 200 μM, with the bidirectional arrow indicating the definition of the peak current, $\Delta I$, as the output signal (and the single-headed arrow showing the direction of the potential sweep); (b) in the absence of G6P and NAD$^+$. The experiment was performed in 10 mM PBS buffer (pH = 7.4) in the presence of 10 mM MgCl$_2$ at ambient temperature (23 ± 2)°C. Potential scan rate: 30 mV s$^{-1}$.

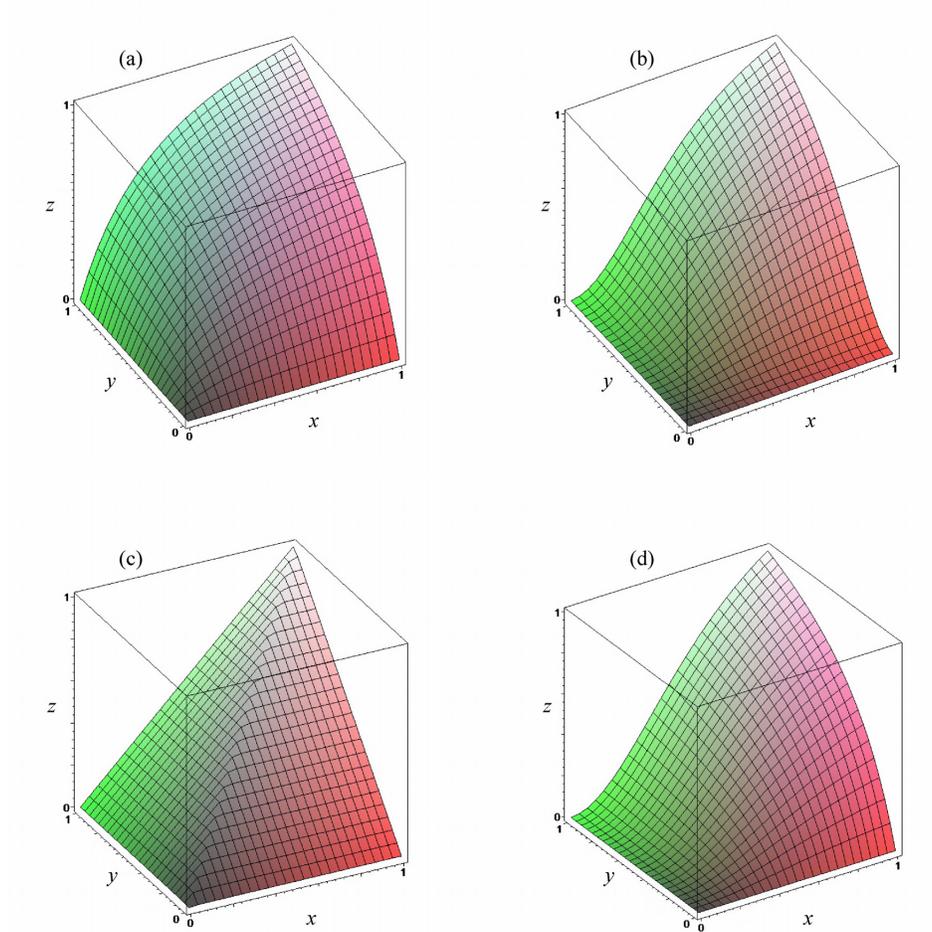

**Figure 3:** Schematic of several types of response surfaces. (a) Typical response surface for biocatalytic reactions. (b) Desirable response surface with "sigmoid" features in both inputs. (c) Response surface that allows elimination of noise amplification without sigmoid behavior. (d) Response surface of the type realized in the present system, "sigmoid" in one of the inputs.

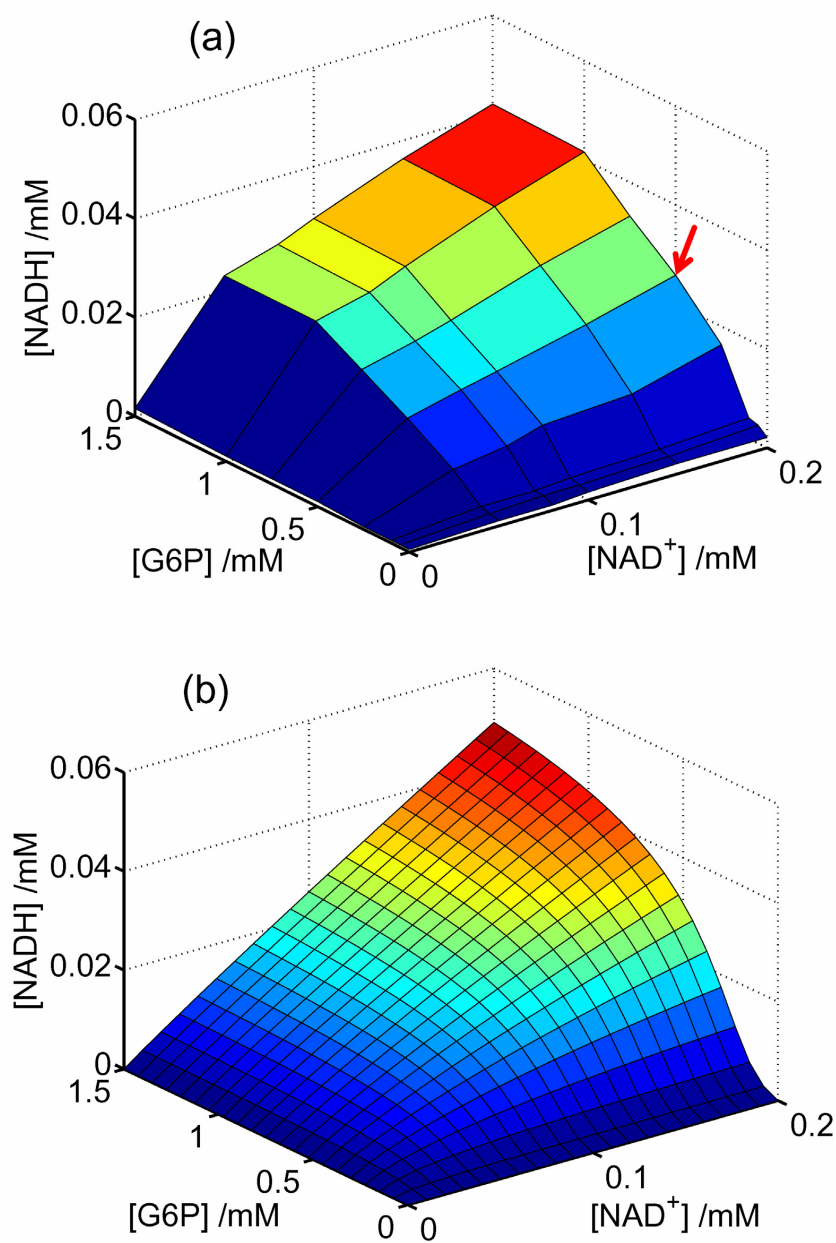

**Figure 4:** (a) Our experimental data, taken at $t_{\text{gate}} = 300\,\text{s}$, illustrating the sigmoid response as a function of the initial input [G6P] value. The arrow marks the values corresponding to the sample voltammogram (a) in Figure 2. (b) Least-squares fit of the response surface, with the fitted parameter values reported in the text.

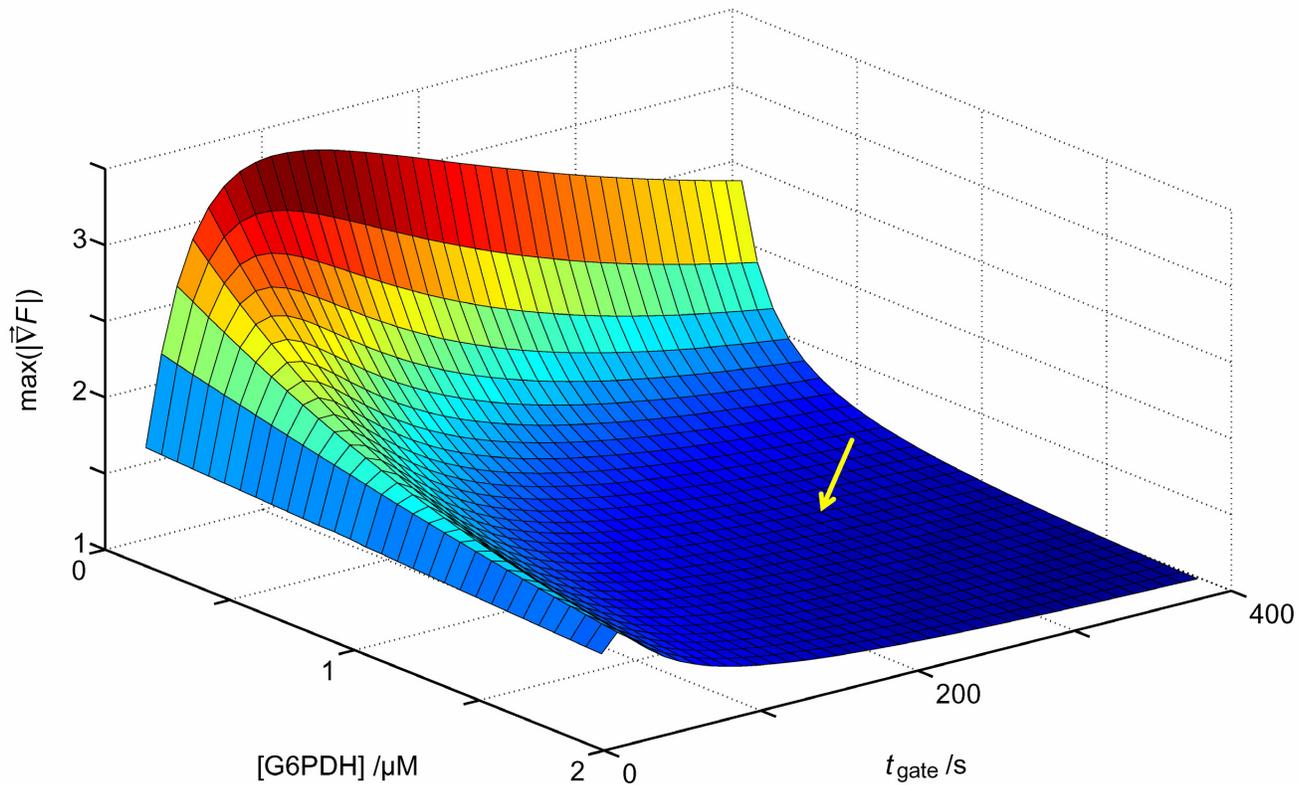

**Figure 5:** Maximum gradient, $|\vec{\nabla} F|_{01}$, of the logic-variable response surface, shown as a function of the reaction time and effective initial enzyme concentration. Our experimental conditions are marked by the arrow. The value of the maximum gradient at this point is 1.16.

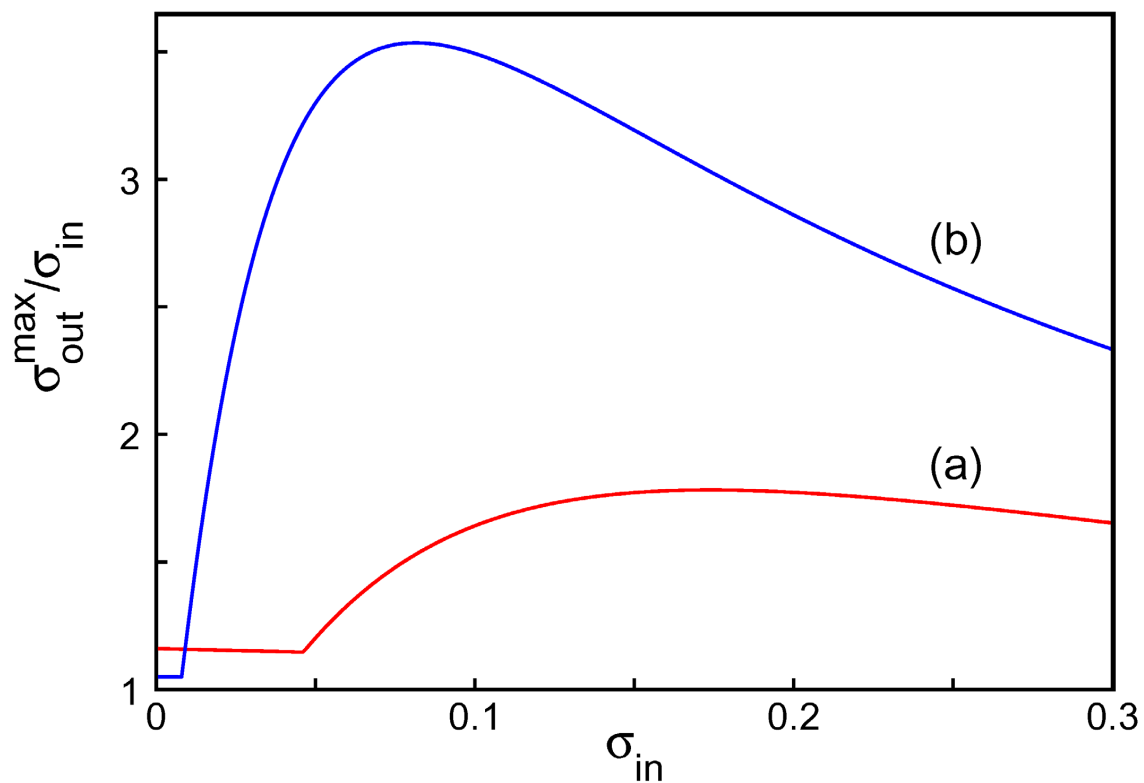

**Figure 6:** Quality measure of the gate response function, $\sigma_{out}^{max}/\sigma_{in}$, vs. the input noise width, $\sigma_{in}$. (a) Calculation for our experimental conditions, marked in Figure 5, with the maximum gradient 1.16. (b) Calculation for the active enzyme concentration 2 μM and time 400 s, i.e., the rightmost point on the surface shown in Figure 5, with the maximum gradient ~1.05.